\documentclass{article}
\usepackage{graphicx}

\begin{document} 

\title{Are we observing black hole spins in jetted Gamma-Ray Bursts?}

\author{Guido Barbiellini$^{1}$, Annalisa Celotti$^{2}$ and Francesco
Longo$^{1}$ \\(1) Department of Physics and INFN, via Valerio 2,
I-34100 Trieste, Italy \\ (2) SISSA, via Beirut 2-4, I-34014 Trieste,
Italy}

\maketitle

\begin{abstract}
We note that if the GRB phenomenon follows from the collapse of a
massive object forming a black hole and a torus accreting into it, the
resulting ejecta can be only related to the mass and angular momentum
that characterize the black hole. The opening angle of the fireball
then corresponds to a black hole spin.  If indeed the fireballs opening
angles are as small as recent estimates reported in the literature
suggest, then GRB would be produced by almost maximally rotating BH.
\end{abstract}

\section{GRB from maximally rotating black holes ?}

The most popular models for the engine of long GRB attribute the
energetics of the prompt emission ($E \sim 10^{52}$ erg) and of the
afterglow, to the energy released by the accretion of a torus onto a
rotating Black Hole (BH) (e.g. Woosley 1993, Vietri \& Stella 1998,
see Meszaros 2002 for a review).

Indeed relativistic numerical simulations and analytical calculations
support the view that the collapse of a rotating massive star can lead
- under rather general assumptions - to the formation of a black hole
and a disc containing about 10 per cent of the mass (Shibata \&
Shapiro 2002, Shapiro \& Shibata 2002).

If the event is catastrophic and not repetitive, the accretion of the
whole of the torus should leave as a final result a BH with mass $M$
and angular momentum $J$, depending on the mass and angular momentum
of the progenitor star.

From general principles (e.g. the no-hair theorem (Misner, Thorne,
Wheeler 1973), the ejecta giving raise to the GRB and afterglow events
can be only related to these quantities, $M$ and $J$, that
characterize the BH.

Within this scenario we then notice that the BH spin is expected to be
directly related to the collimation angle $\theta_{\mathrm jet}$ of
the relativistic fireball responsible for the GRB emission and vice
versa estimated values of $\theta_{\mathrm jet}$ can be translated
into a value of the BH spin $\propto \cos\theta_{\mathrm jet}$. In
particular the results presented by Berger, Kulkarni \& Frail (2003)
correspond to a distribution which suggests that GRB could be produced
by almost maximally rotating BH (see Figure 1).

\begin{figure}
\centering \includegraphics[bb=16 140 580 705, clip=,
width=0.9\textwidth]{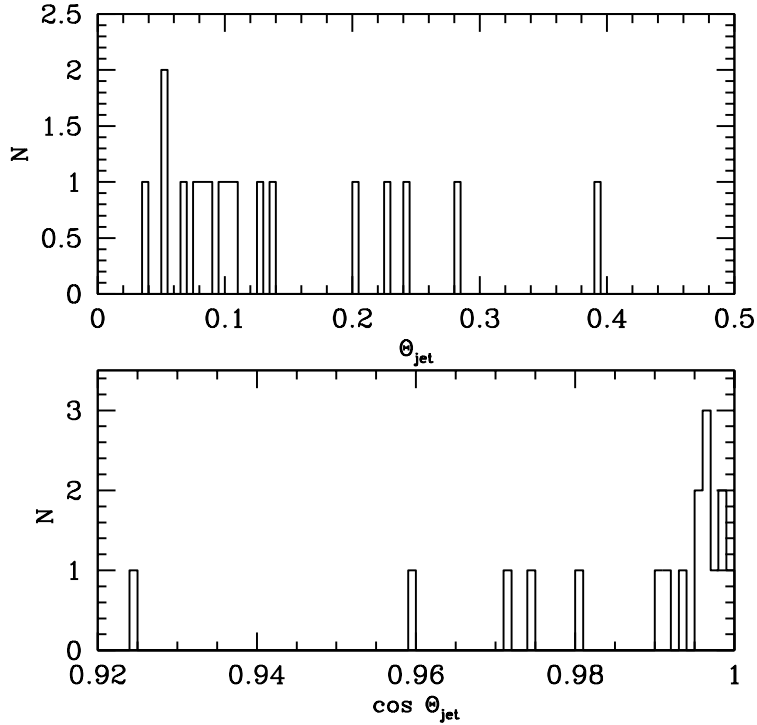}
\caption{Cosine of $\theta_{\mathrm jet}$ reported by Berger, Kulkarni
\& Frail (2003).}
\end{figure}

\section{Conclusions}

Under the general assumption that the energy released in GRB
originates from a torus accreting onto a rotating BH, we note the
collimation angles of the ejecta reflect the distribution of the BH
spin. The reported values (e.g. Berger, Kulkarni \& Frail 2003)
suggest a maximally rotating BH.


\begin{thebibliography}{}

\bibitem{bkf}Berger E., Kulkarni S. R., Frail D. A., 2003, ApJ, submitted
(astro-ph/0301268)

\bibitem{mes02}Meszaros P., 2002, ARA\&A, 40

\bibitem{ss02a}Shapiro S. L., Shibata M., 2002, ApJ, 577, 904

\bibitem{ss02b}Shibata M., Shapiro S. L., 2002, ApJ, 572, L39

\bibitem{vie98}Vietri M., Stella L., 1998, ApJ, 507, L45

\bibitem{mtw}Misner C.W., Thorne K.S., Wheeler J.A., 1973, 
Gravitation (Freeman: San Francisco)

\bibitem{woo93}Woosley S.E., 1993, ApJ, 405, 273

\end{thebibliography}
\end{document}